# Dynamic flows create potentially habitable conditions in Antarctic subglacial lakes

Louis-Alexandre Couston[1,2,3]* and Martin Siegert[4]

Trapped beneath the Antarctic ice sheet lie over 400 subglacial lakes, which are considered to be extreme, isolated, yet viable habitats for microbial life. The physical conditions within subglacial lakes are critical to evaluating how and where life may best exist. Here, we propose that Earth's geothermal flux provides efficient stirring of Antarctic subglacial lake water. We demonstrate that most lakes are in a regime of vigorous turbulent vertical convection, enabling suspension of spherical particulates with diameters up to 36 micrometers. Thus, dynamic conditions support efficient mixing of nutrient- and oxygen-enriched meltwater derived from the overlying ice, which is essential for biome support within the water column. We caution that accreted ice analysis cannot always be used as a proxy for water sampling of lakes beneath a thin (<3.166 kilometers) ice cover, because a stable layer isolates the well-mixed bulk water from the ice-water interface where freezing may occur.

## INTRODUCTION

The Antarctic continent is covered with ice, growing and shrinking over periods of tens to hundreds of thousands of years, since at least the last 14 million years (1). Over 250 hydrologically stable subglacial lakes (in which water inputs are constantly balanced by outputs) trapped between the bed and the ice are known to exist at and close to the ice sheet center (2). They comprise a wide variety of sizes and glaciological and topographic settings (3) and have been hypothesized as potential habitats for the in situ development of microbial organisms (4). Such remote, extreme, and isolated places qualify as analogs to extraterrestrial environments where life may occur, such as the subsurface oceans on Jovian and Saturnian moons (5). A further ~130 hydrologically active lakes, which experience rapid water discharges and large volume changes, exist toward the margin of the ice sheet (6, 7). While these may contain microbial life (8), they are not considered as isolated habitats where microbes can adapt independently over long periods due to the flushing of water in and out of their systems and their potentially ephemeral nature.

The Antarctic ice and bed material carry life's building blocks, with oxygen and minerals held within dust in the former, and minerals trapped inside sediments and bedrock in the latter. Numerical models and radar observations have shown that the ice sheet base above subglacial lakes typically melts where the ice is thickest and freezes where it is thinnest (4, 9, 10). Thus, oxygen and minerals are released at the top of the water column. The rate at which this happens is key to assessing the possibility of having a biome, but remains largely uncertain. Although microbial life is anticipated at the floors of subglacial lakes, where sediments are known to exist (11), dynamic flows and mixing of bottom water within the water column are essential for life to be widespread and detectable, avoiding, for example, anoxic conditions if oxygen-rich surface water is unable to access deeper parts of the lake.

Subglacial lakes are isolated from winds and solar heating but can experience vertical convection flows due to the upward geothermal flux [at a background level of roughly 50 mW/m²; (12)], and horizontal convection flows due to the ubiquitous—albeit variable—tilt of their ice ceiling (about 10 times and in opposite direction to the ice surface slope). Previous work has estimated that velocities of few tenths of a millimeter/second are generally required to suspend sediments in the water column (13). This is of the same order of magnitude as that predicted by ocean modeling for a handful of subglacial lakes, including Lake Vostok (14), Lake Ellsworth (15), and Lake Concordia (9). However, uncertainties are large, and velocities remain unknown for most subglacial lakes, including Lake CECs, which might be the first stable lake to be drilled into in a clean way in the coming years (16). As a result, plans for direct sampling can be helped by establishing hydrological conditions in subglacial lakes, and their variation between lake settings, to recognize where microbial life is most likely to thrive.

Here, we predict the intensity of turbulence and large-scale water circulation for the entire range of stable subglacial lakes found in Antarctica, i.e., with ice cover thicknesses up to 5 km and lake water depths up to 1.5 km (Fig. 1). Thus, our work complements previous studies on convection in lakes at atmospheric pressure (i.e., open) or with thin ice covers (17) and, more specifically, previous efforts that aimed to predict the hydrological conditions of individual subglacial lakes, including Lake Vostok (18). We demonstrate that most subglacial lakes have large supercritical convective parameters, i.e., the geothermal flux is much larger than the minimum critical heat flux required to trigger convective flows, such that they are in a regime of vigorous turbulent convection. We show that vertical convection is as important as horizontal convection and that the convective dynamics vary considerably based on the ice thickness, water depth, and ceiling slope. For simplicity, we restrict our attention to freshwater, because salt concentration is typically low in isolated subglacial lakes (4).

We first calculate the minimum critical heat flux, $F_c$, required to trigger thermally forced vertical convection (Fig. 3) by solving an eigenvalue problem for the local stability of subglacial lakes with a nonlinear equation of state (19). We show that $F_c$ is much smaller than 50 mW/m², which is (approximately) the average geothermal flux, for a wide range of geophysical conditions and conclude that most Antarctic subglacial lakes are unstable to convection. We then demonstrate that most subglacial lakes (Figs. 4 and 5) subject to a geothermal flux of 50 mW/m² experience dynamic flows by applying state-of-the-art scaling laws of classical thermal convection to convection in cold-temperature high-pressure lake environments.

[1]British Antarctic Survey, Cambridge CB3 0ET, UK. [2]Department of Applied Mathematics and Theoretical Physics, University of Cambridge, Cambridge CB3 0WA, UK. [3]Univ Lyon, ENS de Lyon, Univ Claude Bernard, CNRS, Laboratoire de Physique, F-69342 Lyon, France. [4]Grantham Institute and Department of Earth Science and Engineering, Imperial College London, London, SW7 2AZ, UK.
*Corresponding author. Email: louis.couston@ens-lyon.fr

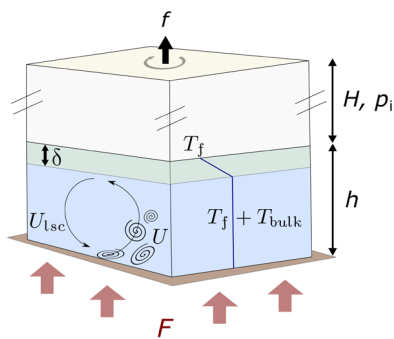

**Fig. 1. Problem schematic.** We provide predictions about the characteristic velocity of the large-scale circulation $U_{lsc}$, the characteristic velocity of turbulent plumes $U$, the thickness $\delta$ of the top stable conductive layer, and the anomalous temperature of the well-mixed bulk $T_{bulk}$ (i.e., in excess of the freezing temperature $T_f$). The problem parameters are the water depth $h$, the ice thickness $H$ (or ice overburden pressure $p_i$), the Coriolis frequency $f$ (due to Earth's rotation), and the geothermal flux $F$.

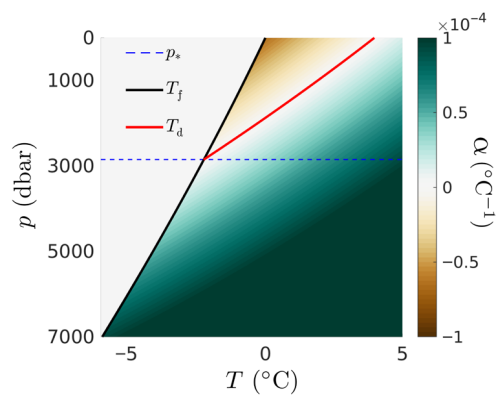

**Fig. 2. Thermal expansion coefficient.** Plot of the thermal expansion coefficient $\alpha$ as a function of $(T, p)$ superimposed with profiles of the temperature of maximum density $T_d$ (red solid line) and freezing temperature $T_f$ (black solid line) with pressure. For small pressures $p < p^*$, with $p^*$ the critical inversion pressure (blue dashed line), $T_d > T_f$ such that there exists a range of temperatures $T_f < T < T_d$ for which $\alpha$ is negative (area appearing with red colors) and water masses become anomalously denser with increasing temperatures. For $p > p^*$ and $T > T_d$, or $p \geq p^*$, the water becomes monotonically lighter as temperature increases, which is the typical behavior of most fluids.

## RESULTS

Subglacial lakes can experience dynamic vertical convection flows [also known as Rayleigh-Bénard (RB) convection (20)], because the lake's deepest waters, heated by Earth's geothermal flux, are generally buoyant and will tend to rise through and mix with the rest of the water column. How far up and how quickly bottom water masses rise depend on the geothermal flux, $F$, the water depth, $h$, and the ice cover thickness, $H$ (or ice overburden pressure $p_i$). Convection in subglacial lakes is complex, because the thermal expansion coefficient of freshwater $\alpha(p, T)$, which indicates how fluid parcels contract or expand with changes in temperature (i.e., are buoyant), depends on water pressure $p > p_i$ and the temperature $T$ itself (21). For relatively thick ice cover, i.e., $H \geq H^* = 3166$ m, the thermal expansion coefficient is always positive (i.e., the density decreases with temperature) and increases with pressure and temperature, such that convection becomes more vigorous as $F$, $h$, and $H$ increase. For ice covers less than the critical ice depth $H^*$, or ice pressure $p_i < p^* = 2848$ dbar (which we refer to as the critical ice pressure), however, the thermal expansion coefficient changes sign with temperature, such that density does not simply decrease with temperature but becomes a nonlinear and non-monotonic function of $T$. Specifically, as is shown in Fig. 2, for $H < H^*$, $\alpha$ increases with temperature but is first negative for $T_f(p_i) \leq T < T_d(p)$ (close to the ice ceiling), where $T_f$ is the freezing temperature and $T_d$ is the temperature of maximum density, before becoming positive at higher temperature. Having $\alpha < 0$ close to the ice ceiling for $H < H^*$ means that the density stratification is always stable at the top of the lake and that the bottom layer is buoyant only if the bottom temperature exceeds $T_d$, i.e., such that the density stratification is top heavy near the bottom. Having $\alpha > 0$ on the bottom boundary is a necessary condition for deep water masses to be buoyant but, however, is not sufficient for convection to set in. The geothermal flux must be also larger than the adiabatic heat flux and adequate to sustain a buoyancy force that can overcome viscous dissipation and thermal diffusion.

Here, we estimate the minimum critical heat flux that overcomes dissipation effects and permits convection in subglacial lakes from a stability analysis of the perturbation equations for a water column subject to geothermal heating and the Coriolis force due to Earth's rotation. We consider a realistic nonlinear equation of state for freshwater using the Thermodynamic Equation of Seawater 2010 (TEOS-10) toolbox (19) and the Coriolis frequency at 80°S. We take into account the adiabatic temperature gradient by including compressibility effects in the energy equation. We perform the calculations for a wide range of ice thicknesses and water depths up to 20 m. For water depth, $h > 20$ m, the eigenvalue problem becomes too difficult to solve, so we use scaling laws that are either conservative or inferred from classical RB convection results in the limit of rapid rotation (22).

Figure 3A shows the minimum critical heat flux $F_c$ that permits vertical convection for a wide range of ice pressures and water depths relevant to Antarctic subglacial lakes. $F_c$ is large at small pressure and small water depth (top left corner) but then decreases with $h$ and $p_i$ in most of the parameter space. We find $F_c < 50$ mW/m² (as shown by the black isocontour labeled "50"), which is a typical background value for Earth's geothermal flux around Antarctica, in most of the parameter space. We predict that Lake CECs and South Pole Lake (SPL) are unstable to vertical convection if subject to a 50 mW/m² flux, i.e., their critical heat flux $F_c$ is less than 50 mW/m², despite having relatively thin ice covers $H < H^*$ (shown by the gray dashed line). Here, we have centered the vertical axis of Fig. 3A on the critical pressure $p^*$ by using the shifted ice pressure variable $p_i - p^*$ and a symmetric logarithmic scale. As a result, the transition from a fully convective water column (for $p_i > p^*$) to a convective water column with a stably stratified upper layer (for $p_i \leq p^*$) is smooth even though $F_c$ increases rapidly as $p_i$ decreases below $p^*$. Figure 3B shows that subglacial lakes reported in the last inventory (2) have ice cover thicknesses almost equally distributed on either side of $H^*$. Thus, the $p_i = p^*$ isobar, which separates lakes that are fully convecting from lakes that are only partially convecting, is important not only for the theoretical calculation of $F_c$ but also in practice. Almost half of the subglacial lakes (with $p_i < p^*$) may be expected to have a top layer that is stable, although possibly modified by the dynamics near the ice ceiling and overshooting convection. We remark that $F_c$ is constrained primarily by the condition of having $\alpha > 0$ on the bottom boundary for small $p_i$ and $h$ (top left corner) and by the condition that it must exceeds the adiabatic flux for large $p_i$ and $h$ (bottom right corner). In between, viscous dissipation and thermal diffusion dominate the calculation of $F_c$. Note that our prediction of $F_c$ for small $p_i$ and large $h$ is conservative and may overestimate the true $F_c$. We provide further details about the calculation of $F_c$ in Materials and Methods and in the Supplementary Materials.

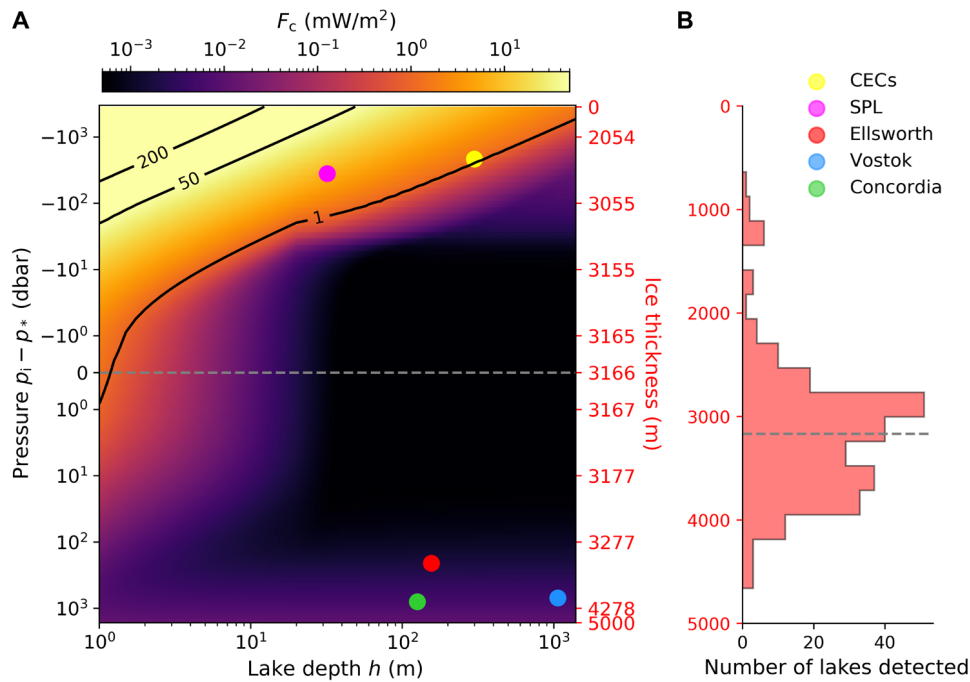

**Fig. 3. Critical heat flux.** (**A**) Minimum heat flux required to trigger vertical convection in subglacial lakes as a function of lake depth (bottom axis) and ice overburden pressure (left axis) or ice thickness (right axis). Solid lines are isocontours in mW/m² of required heat flux, while filled circles highlight the positions of five well-known lakes in parameter space (see legend to the right). (**B**) Ice thickness distribution of isolated subglacial lakes from the last published inventory (2). The dashed gray lines highlight the critical thickness $H^*$.

For a geothermal flux $F$ greater than the critical heat flux $F_c$, it is of interest to know whether the convective instability results in high or low velocities. In general, estimates of flow velocities require dedicated simulations or laboratory experiments. For the case of turbulent vertical convection, however, various scientific communities have proposed predictive laws of hydrodynamic variables based on control parameters, which can be used as leading-order estimates for turbulence intensity in convective subglacial lakes [see (20) and references therein]. The canonical problem of natural convection relevant to our work is known as rotating RB convection and has applications in many different fields, including geophysics (23), astrophysics, and engineering (20). Hydrodynamic variables such as turbulent flow velocities, large-scale flow velocities, and temperature fluctuations are predicted on the basis of the value of the Rayleigh number $Ra$ of the system, which is a dimensionless measure of the available convective energy; the Prandtl number $Pr$, which compares viscous dissipation to thermal diffusion; and the Ekman number $Ek$, which compares viscous dissipation to the Coriolis force. The Rayleigh number, Prandtl number, and Ekman number for subglacial lakes in Antarctica can be written as

$$Ra_F = \frac{g \alpha_{eff} F h_{eff}^4}{\nu \kappa k}, \; Pr = \frac{\nu}{\kappa}, \; Ek = \frac{\nu}{|f| h_{eff}^2} \quad (1)$$

where $g$ is the gravitational acceleration, $\alpha_{eff}$ is the characteristic thermal expansion coefficient, $h_{eff}$ is the effective water depth where convection occurs, $\nu$ is the kinematic viscosity, $\kappa$ is the thermal diffusivity, $k$ is the thermal conductivity, and $f$ is the Coriolis frequency (note that we use $|f|$ in our definition of $Ek > 0$, since $f < 0$ in the Southern Hemisphere). The subscript $F$ of $Ra_F$ means that the Rayleigh number of subglacial lakes is a flux-based Rayleigh number, since it is based on a prescribed geothermal flux rather than a prescribed temperature difference, which is more common in idealized studies of natural convection (20). Note that we neglect compressibility effects for the prediction of variables in the turbulent regime because Earth's geothermal flux is several orders of magnitude larger than the adiabatic flux.

In the context of subglacial lakes, the geothermal flux is sufficiently large that the lake water is in a fully turbulent state that is almost not affected by rotation, i.e., $F \gg F_c$, and the effect of rotation is weak [see the Supplementary Materials and (24)]. Thus, here we use scaling laws derived for fully turbulent nonrotating convection to make predictions about hydrodynamic variables. The variables of interest are the thickness of the conductive layer near the ice ceiling $\delta$, the anomalous temperature of the well-mixed bulk $T_{bulk}$ (in excess of $T_f$), the characteristic turbulent flow velocity $U$, and the length scale of turbulence $\ell$, which represents the typical distance between thermal plumes (Fig. 1). We assume a geothermal flux of $F = 50$ mW/m² throughout and use scaling laws derived from numerical simulations (25) as well as scaling laws inferred from the Grossmann-Lohse (GL) unifying theory of RB convection, which is based on theoretical arguments (26).

We first show in Fig. 4 (A to D) the results for $\delta$ and $T_{bulk}$ based on the scaling laws derived in (25) for a wide range of ice thicknesses and water depths. The thickness of the conductive layer near the ice ceiling is almost independent of lake depth but varies substantially with ice pressure (Fig. 4, A and B). For thin ice cover, the top conductive layer consists of a layer with a stable density stratification attached to the ice ceiling (where $\alpha < 0$), which can be several meters thick (thickness $\delta_S$), and a turbulent transition layer (just above the convective bulk), which is typically on the order of a few centimeters or smaller (thickness $\delta_t$), i.e., $\delta = \delta_S + \delta_t$. For ice thickness, $H < 2000$ m, $\delta_S$ is between 10 and 40 m. The stable layer thickness $\delta_S$ decreases with $H$ and vanishes for $H > H^*$ (since $\alpha > 0$ everywhere in this case), such that the full conductive layer is limited—for a thick ice cover—to a turbulent boundary layer attached to the ice ceiling, which is small. In all cases, the top stable layer transfers heat by conduction only. Thus, the temperature increases linearly from $T_i(p_i)$ near the ice to $T_i(p_i) + \delta F/k$ at the bottom of the stratified layer. The anomalous temperature of the well-mixed convective bulk (above freezing) can then be approximated as $T_{bulk} = \delta F/k$ (Fig. 4, C and D), and hence shows similar trends as δ. The white area on the top left corners of Fig. 4 (A and C) highlights subglacial lakes that are stable because the thermal expansion coefficient is negative everywhere in the water column. The filled squares in Fig. 4 (B and D) show δ and $T_{bulk}$ based on scaling laws derived in (26) (labeled "GL"). There is a good agreement between predictions based on the scaling laws in (25) (shown by circles) and (26).

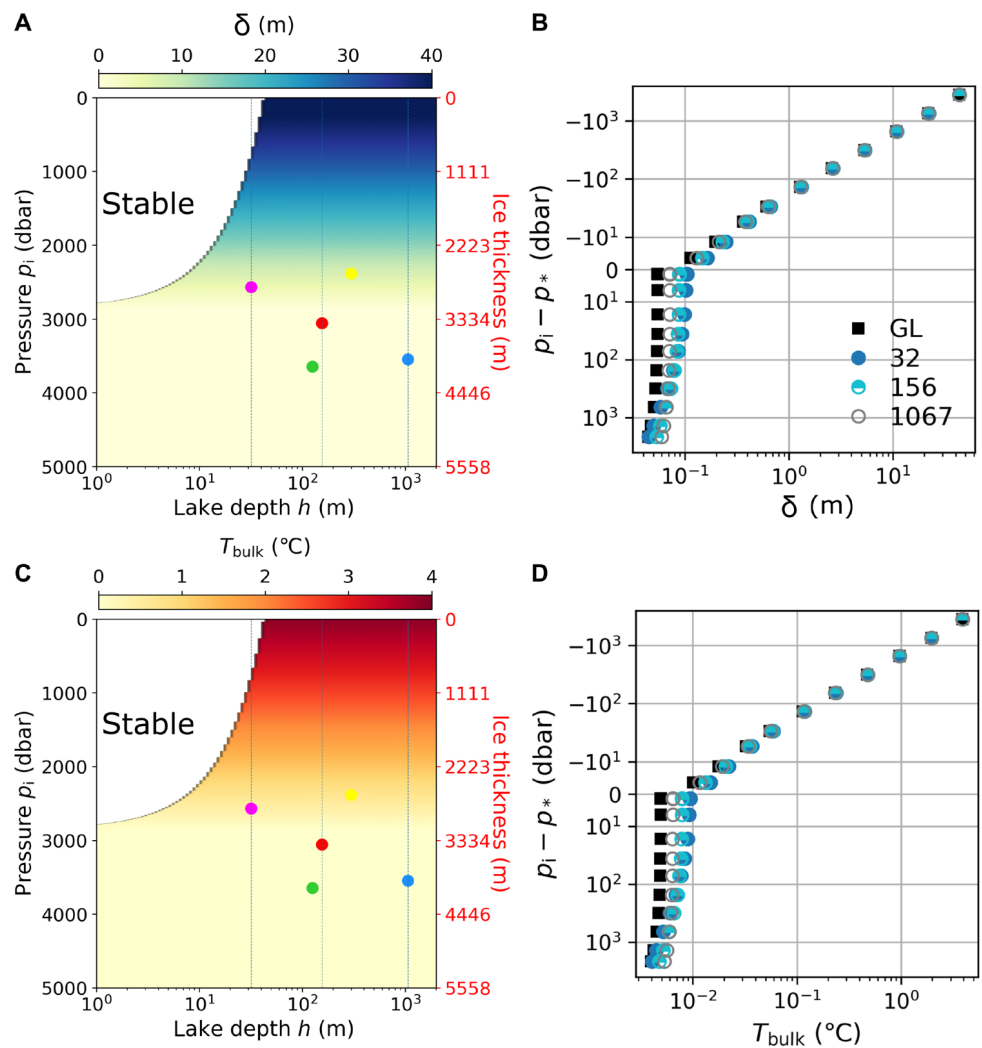

**Fig. 4. Conductive layer thickness and anomalous bulk temperature.** (**A** and **B**) Thickness δ of the conductive stratified layer at the top of subglacial lakes assuming a geothermal flux of 50 mW/m². (A) δ as a function of lake depth (bottom axis) and ice pressure (left axis). (B) δ as a function of ice pressure only for selected lake depths of 32, 156, and 1067 m [shown by vertical lines in (A)]. GL refers to results obtained with the GL theory and $h = 1067$ m. (**C** and **D**) Same as (A) and (B) but for the anomalous bulk temperature $T_{bulk}$ (above $T_f$) of the well-mixed convective layer.

Figure 5 shows the predicted lake velocity $U$ and horizontal length scale $\ell$ based on previously derived scaling laws (25). Figure 5 (A and B) shows that $U$ is almost independent of ice pressure but increases with lake depth, up to about 1 cm/s for $h = 1500$ m. Figure 5 (C and D) shows that $\ell$ increases slowly with lake depth and remains on the order of 1 m for all ice pressures. Both $U$ and $\ell$ appear discontinuous at $p_i = p^*$ because the effective thermal expansion coefficient $\alpha_{eff}$, which we estimate conservatively (cf. Materials and Methods) and use in Eq. 1 for $Ra_F$, decreases rapidly across the $p^*$ isobar for small water depths. For instance, $\alpha_{eff}$ decreases from $3 \times 10^{-6}$ °C$^{-1}$ to $3 \times 10^{-7}$ °C$^{-1}$ between $p_i = p^* + 100$ dbar and $p_i = p^*$ for $h = 10$ m. We expect that the discontinuity would become less sharp but would not completely disappear upon relaxing our conservative approximation for $\alpha_{eff}$, because $\alpha$ will always be (overall) much smaller in lakes with a thin ice cover ($p_i \leq p^*$) than in lakes with a thick ice cover ($p_i > p^*$). We also show in Fig. 5B the prediction for the characteristic velocity $U_{lsc}$ based on the scaling laws derived in (26) (labeled GL) for an ice thickness $H = 3945$ m (filled squares), which is the ice thickness above Lake Vostok. $U_{lsc}$ is smaller than $U$ (shown by the filled circles) by up to a factor 5 because the GL theory focuses on the velocity of the large-scale circulation, while $U$ is the characteristic root-mean-square velocity of turbulent plumes (25), which is likely to be faster than the mean large-scale flow. Figure 5B also shows a prediction for the horizontal velocity $V_{hc}$ of the baroclinic flow expected along a sloped ice-water interface, using scaling laws inferred from recent results on horizontal convection (27). We show the prediction for $V_{hc}$ assuming two different ice-water interface slopes, i.e., $s = 10^{-3}$ and $s = 10^{-2}$, the steepest slope resulting in the largest horizontal velocity due to the increased temperature gradient along the ice ceiling. The horizontal velocity of the baroclinic flow is of the same order (for a steep slope, $s = 10^{-2}$) or one order of magnitude smaller (for a moderate slope, $s = 10^{-3}$) than the large-scale velocity of vertical convection ($U_{lsc}$).

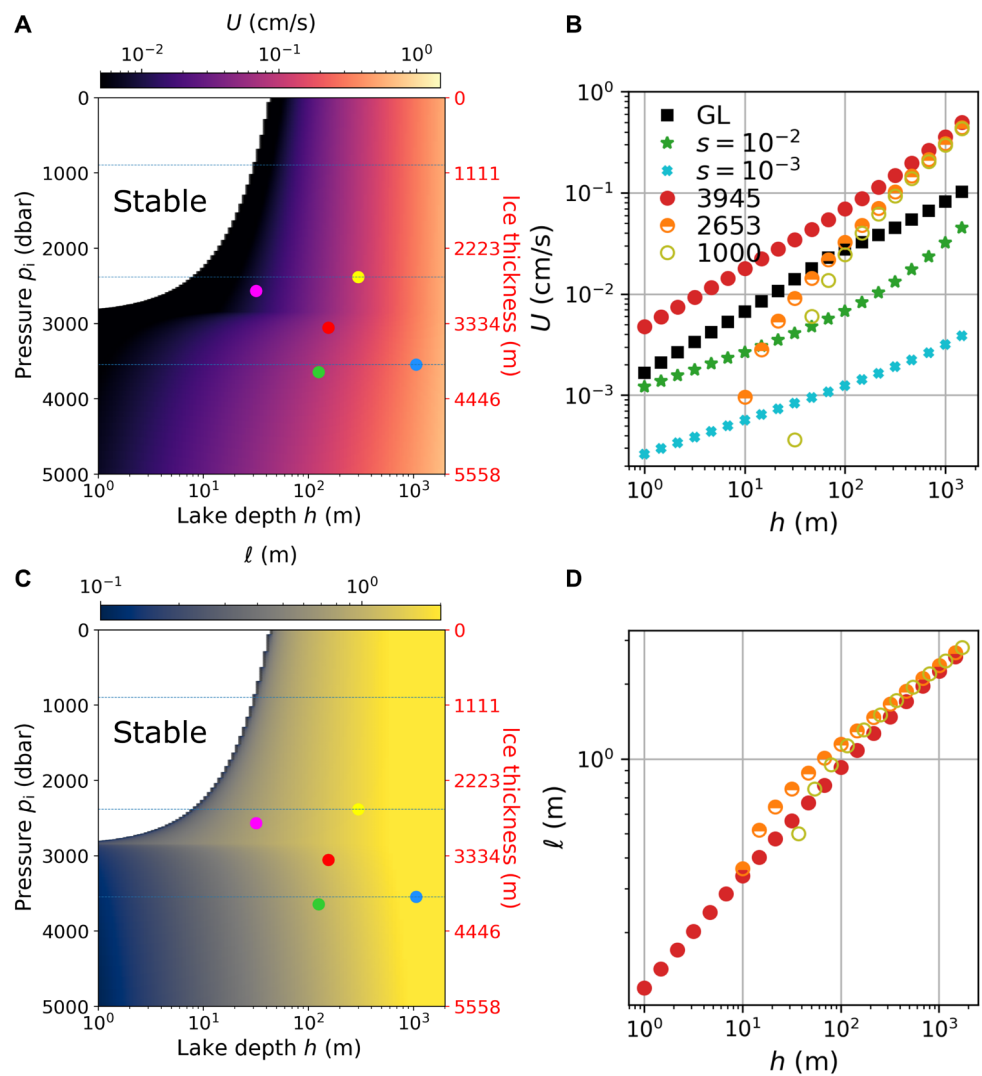

**Fig. 5. Characteristic turbulent flow velocity and length scale.** (**A** and **C**) Same as Fig. 4 (A and C), but for (A) the turbulent flow (plume) velocity $U$ and (C) the characteristic length scale $\ell$ in the convective layer. (**B** and **D**) Turbulent flow velocity $U$ and length scale $\ell$ as functions of lake depth only, for selected ice thicknesses $H = 3945$, 2653, and 1000 m [shown by horizontal lines in (A) and (C)]. GL refers to the GL predictions for the large-scale velocity $U_{lsc}$ of vertical convection for $H = 3945$ m (shown by filled black squares). In (B), we also show a prediction for the horizontal velocity $V_{hc}$ of the baroclinic flow along a tilted ice-water interface, assuming either a steep slope $s = 10^{-2}$ (green stars) or a moderate slope $s = 10^{-3}$ (tilted blue crosses).

## DISCUSSION

We have demonstrated that the critical heat flux leading to vertical convection in subglacial lakes is much less than 50 mW/m² for a broad range of ice overburden pressures and water depths (Fig. 3). Thus, it should be considered that most—if not all—Antarctic subglacial lakes are dynamic hydrologic environments. We expect that the same conclusion holds for isolated subglacial lakes in Greenland and elsewhere in the solar system (5, 28). We note that our prediction of the critical heat flux $F_c$ is conservative for large water depth and small ice pressure (see the Supplementary Materials). Also, estimates of $F_c$ exceeding 50 mW/m$^2$ (as is the case for a lake 20 m deep and under 1 km of ice), hence suggesting stable subglacial lakes, could be verified qualitatively through direct sampling and revised if measurements demonstrate a dynamic environment.

Vertical convection in subglacial lakes is different from vertical convection in the canonical RB problem, mainly because the thermal expansion coefficient ($\alpha$) of freshwater in subglacial lakes changes with pressure and temperature (Fig. 2), while it is typically constant in RB studies. $\alpha$ is negative at low pressures and low temperatures, such that a layer of stable density stratification exists at the top of subglacial lakes beneath a thin ice cover ($H < H^*$). Here, we have used state-of-the-art scaling laws of RB convection and took into account the variability of the thermal expansion coefficient to make predictions about the thickness of the stable layer near the ice ceiling ($\delta$), the anomalous temperature of the well-mixed turbulent bulk ($T_{bulk}$), the characteristic velocity of plumes ($U$), the characteristic velocity of the large-scale circulation ($U_{lsc}$), and the characteristic distance between plumes ($\ell$). For completeness, we have also used state-of-the-art scaling laws of horizontal convection to make predictions about the typical horizontal velocity ($V_{hc}$) of the baroclinic flow that develops along a tilted ice-water interface.

The predictions for the different hydrodynamic variables are shown in Figs. 4 and 5. The key result of Fig. 4 is that subglacial lakes with a thin ice cover have an upper conductive layer several meters thick and a warm turbulent bulk (up to 1 K above freezing), whereas subglacial lakes with a thick ice cover have a thin conductive layer (centimeter scale) and a cold turbulent bulk beneath (0.01 K or less above freezing). The key result of Fig. 5 is that subglacial lakes deeper than 100 m experience substantial flow velocities, specifically $U \approx 4$ mm/s and $U_{lsc} \approx 1$ mm/s for a 1-km deep lake. These vertical velocities are larger than the horizontal velocity associated with the baroclinic flow along a tilted ice-water interface, even if the ice slope is as large as $s = 10^{-2}$. The ratio $U_{lsc}/V_{hc}$ is not much larger than 1 for steep slopes. However, if we assume that the vertical velocity of the baroclinic flow scales like $\sim V_{hc}h/L$ with $L \gg h$ the horizontal length of the lake, then $U_{lsc}L/(V_{hc}h) \gg 1$. Thus, geothermal heating is a key factor—if not the dominant one—controlling hydrological conditions in Antarctic subglacial lakes. We provide predictions of flow properties for five well-studied subglacial lakes in Table 1.

Our analysis assumes that vertical convection and horizontal convection are decoupled. This limitation comes from the fact that numerical simulations and laboratory experiments tackling both dynamics simultaneously (either in a realistic or idealized setting) are lacking. A handful of coarse numerical models have provided some insights into the large-scale circulation of select subglacial lakes (9, 10, 14), but they rely on approximations (including parameterized turbulent processes) and are too expensive to run to allow the derivation of scaling laws of combined vertical and horizontal convection. There has been only one attempt so far at a laboratory analog of subglacial lake dynamics (29) in which the combined convective dynamics, dominated by rotation and taking the form of columnar vortex structures, was observed. The possibility to have vortices extending throughout the entire water column at full scale is an open question, which would be worth exploring.

Future work should also consider investigating the importance of the coupling dynamics between the lake circulation and melting/freezing processes along the ice-water interface. For a flat ice ceiling, we may expect that melting and freezing patterns emerge where vertical convection drives upwelling and downwelling, respectively. Such patterns would be separated by a distance equal to the lake water depth, which is the characteristic length scale of the large-scale circulation of vertical convection. For a tilted ice roof, state-of-the-art numerical models of subglacial lake dynamics typically predict that melting occurs where the ice is thickest (9, 10). However, vertical convection is not well represented in these models such that uncertainties are large regarding melting patterns and the back reaction of melting and freezing processes on the underlying lake dynamics. For instance, melting induced by the baroclinic flow may intensify local vertical convection if the melt water is dense. The possibility that topographical features emerge because of variable melt rates along the ice-water interface and influence the long-term flow dynamics is another interesting point that has yet to be addressed.

Melting and freezing occur as a result of heterogeneous heat fluxes along the ice-water interface driven by the lake water circulation. Melting of the ice ceiling into the lake releases oxygen and minerals trapped in dust particulates, and sediments can be incorporated into the lake from upstream (30). An important question is: What happens to particulates released from the lake roof and how are they dispersed by the lake circulation? Particulates in subglacial lakes can most likely be considered as passive tracers because (i) their characteristic spherical radii, which are in the range 1 μm $< r <$ 100 μm, are much smaller than the Kolmogorov length scale, which is $\eta = h/Re^{3/4} \approx 1$ cm (20) for a typical water depth of $h = 100$ m; (ii) their density $\rho_s$ is larger but of the same order as the density of water, i.e., $\rho_s \approx 3\rho_0$ with $\rho_0 \approx 999$ kg/m$^3$ the mean density of water; and (iii) particulates' loading is expected to be dilute (13). In a quiescent fluid, sub-Kolmogorov particulates settle by gravity with speed $W = 2gr^2(\rho_s - \rho_0)/9\eta$, assuming a linear Stokes drag, with $\eta = 0.0017$ m$^2$ s$^{-1}$ the dynamic viscosity of water. In a convecting fluid, particulates can either settle with a similar velocity or stay suspended provided that the large-scale circulation is upward and has velocity $U_{lsc} > W$, where, here, $U_{lsc}$ is estimated from the GL theory. We report in the last column of Table 1 twice the maximum radius $r_{max}$ of particulates maintained in suspension by the large-scale flow, i.e., such that $W(r_{max}) = U_{lsc}$. We find $2r_{max} > 7.8$ μm in all cases ($2r_{max} > 20$ μm in all lakes but SPL), which means that a broad range of particulates as observed in Vostok's accreted ice, and qualifying as "fine silt," may be suspended in all five lakes. For Lake Vostok, it may be noted that we predict a maximum radius (36 μm) larger than that (23 μm) reported by (13) (and for nonspherical particles, such as micas, the longest axis may be even larger). This difference arises because we calculate larger flow velocities in the lake. The flow velocities and suspended particulates, which we predict for Lake Vostok, would certainly be observable by direct measurements.

**Table 1. Properties and expected characteristics of five Antarctic subglacial lakes.** The last column is the predicted maximum diameter of particulates maintained in suspension in the mixed bulk by the large-scale circulation of vertical convection (see Discussion section). Geophysical characteristics are obtained from (2, 9, 10, 16, 57, 58), while flow conditions are derived from scaling laws discussed in the Results section of the main text and described in detail in the sections, "Scaling laws for nonrotating vertical convection" and "Scaling laws for rotating horizontal convection," in the Materials and Methods. Ice drop refers to the difference in ice thickness above the lake due to the mean slope of the ice-water interface.

| | Ice thickness (m) | Ice drop (m) | Lake length (km) | Water depth (m) | $\delta$ (m) | $T_{bulk}$ (K) | $\ell$ (m) | $U$ (mm/s) | $U_{lsc}$ (mm/s) | $V_{hc}$ (mm/s) | $2r_{max}$ (μm) |
|---|---|---|---|---|---|---|---|---|---|---|---|
| CECs | 2653 | 159 | 10.35 | 300 | 7.7 | 0.69 | 1.6 | 0.97 | 0.32 | 0.041 | 22 |
| SPL | 2857 | 30 | 10 | 32 | 4.7 | 0.42 | 0.8 | 0.10 | 0.04 | 0.010 | 7.8 |
| Ellsworth | 3400 | 300 | 10 | 156 | 0.077 | 0.0069 | 1.2 | 0.69 | 0.26 | 0.066 | 20 |
| Vostok | 3945 | 600 | 280 | 1067 | 0.066 | 0.0059 | 2.3 | 3.80 | 0.85 | 0.066 | 36 |
| Concordia | 4055 | 168 | 45 | 126 | 0.063 | 0.0056 | 1.0 | 0.83 | 0.31 | 0.044 | 22 |

In addition to the large-scale circulation, subglacial lakes experience fast and turbulent motions that can lift sediments from the bed and oppose particulates' settling by dispersing them. The mean vertical distribution of small particulates with low inertia can be approximated by an advection-diffusion equation. The steady-state distribution in such a model is an upward-decaying exponential for the concentration of particulates $n(z) \sim e^{-zW/D}$ in the bulk, which shows that increased turbulence increases the particulates' concentration by raising the background effective diffusivity, which we denote by $D$, in the water column. The background effective diffusivity in the open ocean is well documented and typically ranges from $D = 10^{-5}$ m$^2$ s$^{-1}$ to $D = 10^{-4}$ m$^2$ s$^{-1}$ (31). For a particulate with radius 4 μm and settling velocity $W = 0.04$ mm s$^{-1}$, the corresponding $e$-folding decay length scale ranges from 25 cm to 2.5 m. The effective diffusivity in subglacial lakes is unknown but may be estimated from our predictions for the characteristic velocity $U$ and length scale $\ell$ of plumes as $D \sim U\ell$. For most subglacial lakes, we predict 0.1 mm/s $< U <$ 1 cm/s and $\ell \sim 1$ m (Fig. 5), such that $D \sim U\ell \sim 10^{-4} - 10^{-2}$ m$^2$ s$^{-1}$ and the $e$-folding decay length scale goes from 2.5 to 250 m. Whether an effective diffusivity based on the velocity and length scale of plumes is more applicable than an effective diffusivity typical of the open ocean is an open question. The effective diffusivity based on the properties of plumes is most likely an upper bound, since plumes are intermittent. Thus, we might expect that the mean concentration of particulates, i.e., uniform in space and time, is controlled by a weak diffusivity ($\sim 10^{-4}$) and decays by at least one order of magnitude every 10 m. This means that future explorations limited to sampling in the bulk of the lake would have to rely on intermittent plumes and local upwelling of the large-scale circulation to bring particulates upward. The mean number $N$ of particulates in the water column, or turbidity, is key to fully assessing the habitability of subglacial lakes, in addition to the concentration of oxygen molecules derived from the ice above (11). Our calculations demonstrate that mixing of subglacial lake water is highly likely and would encourage dispersion of oxygen-rich water throughout the water column and down to the lake floor sediments, where microbial life is likely to be most abundant. A comparison of predictions for $N$ based on advection-diffusion models as well as inferred from particulates' concentration in basal and accreted ice, as already done for Lake Vostok (32, 33), will be key to assessing the robustness of the hydrological conditions predicted in this paper and of future particulate distribution models.

This paper provides predictions for flow velocities (0.01 to 1 cm/s), turbulent length scales (1 m), top stable layer thickness (0.01 to 10 m), temperature fluctuations (0.001 to 1 K), and the radius of particulates suspended in the water column (1 to 40 μm) due to vertical convection in Antarctic subglacial lakes. Those predictions will be verifiable by future explorations sampling lake waters and sediments using, e.g., conductivity, temperature, and depth (CTD) profilers, such as envisioned for Lake CECs and as was initially planned for Lake Ellsworth (34). To date, planning for the exploration of Lake Vostok has hinged on the analysis of accreted ice from the lake's water in ice cores (32, 33). Our work shows that such an approach might prove inappropriate for lakes with ice covers thinner than $H^* = 3166$ m, such as Lake CECs, since, in this case, a thick meter-scale stable layer at the top of the water column prevents the upwelling of deep water and its freezing at the ice-water interface. It also means that sampling from Lake CECs should not take place at and close to the ice-water interface; instead, we predict essential measurements are required at least 1 m below the surface of the lake and likely along the entire water column.

We remark that having a stable density stratification at the top of the water column does not imply a completely quiescent environment adjacent to the ice ceiling (even if flat). Internal gravity waves (35) generated by penetrative convection (36) can propagate within the stable layer and affect particulate settling (37). How much energy is transferred from convective motions to internal waves depends on the ratio of the buoyancy frequency of the stable layer, $f_S$, to the convective frequency, $f_c$. For a subglacial lake, such as Lake CECs, we have $f_S = \sqrt{-g\,\rho_0^{-1}\,d\rho/dz} \approx 1$ min$^{-1}$ and $f_c \sim U_{lsc}/h \sim 0.1$ day$^{-1}$. Thus, $f_S \gg f_c$, such that convection is unlikely to penetrate far into the stratified layer and internal wave generation is weak (although this prediction neglects the possible influence of horizontal convection) (36). Nevertheless, it would be worth investigating whether internal waves in subglacial lakes can promote melting or freezing at the ice-water interface. Last, an analysis similar to the one developed in this work could be implemented for predicting dynamic conditions in icy moons in the solar system, where deep subsurface oceans exist and have attracted attention as potential habitats for extraterrestrial life (38).

## MATERIALS AND METHODS

### Equation of state

We use the TEOS-10 toolbox in MATLAB (19) to estimate values for (i) the density of water, $\rho^e(T, p, S)$, as a function of in situ temperature $T$, water pressure $p$, and absolute salinity $S$; (ii) the freezing temperature, $T^e_f(p_i, S)$, as a function of $p_i$ (the pressure at the ice-water interface) and $S$; and (iii) the temperature of maximum density, $T^e_d(p, S)$, as a function of $p$ and $S$. Here, we restrict our attention to freshwater as opposed to seawater, i.e., we set $S = 0$, such that variables do not depend on $S$. We use superscript e to denote exact quantities, and we call the water pressure $p$ the pressure for short, which is the absolute pressure minus atmospheric pressure $p_a = 10.1325$ dbar. Note that $p$ is the full water pressure, which includes pressure contributions from the ice cover, such that $p > p_i$, with $p_i$ the ice overburden pressure. The ice pressure is related to the ice thickness through $p_i = \rho_i g H/10^4$, with $p_i$ in decibars and $H$ in meters, and we assume a mean ice density $\rho_i = 917$ kg/m$^3$. Unless stated otherwise, all variables use SI units except temperature variables, which are in degrees Celsius (°C) and pressure variables, which are in decibars (dbar), since °C and dbar are standard units in physical oceanography.

For simplicity, we derive explicit, approximate expressions for $\rho$, $T_f$, and $T_d$ from the TEOS-10 exact values. The freezing temperature and the temperature of maximum density can be well approximated by quadratic polynomials in $p_i$ and $p$, respectively. For $0 < p$, $p_i < 10^4$ dbar, we find that the best-fit polynomials (with $p$, $p_i$ in dbar)

$$T_f = 4.7184 \times 10^{-3} - 7.4584 \times 10^{-4} p_i - 1.4999 \times 10^{-8} p_i^2 \quad (2)$$

$$T_d = 3.9795 - 2.0059 \times 10^{-3} p - 6.2511 \times 10^{-8} p^2 \quad (3)$$

approximate $T_f^e$ and $T_d^e$ to within 0.002 K. We approximate the density of water by a quadratic bivariate polynomial, which is maximum at $T = T_d(p)$. For $0 < p < 10^4$ dbar and $T_f < T < T_f + 15$ K, we find that the best-fit bivariate polynomial (with $p$ in dbar)

$$\rho = \rho_0 + \rho_1(p) + C(p)\left[T - T_d(p)\right]^2 \quad (4)$$

with

$$\rho_0 = 9.9999 \times 10^2, \rho_1 = 4.9195 \times 10^{-3} p - 1.4372 \times 10^{-8} p^2$$
$$C = -7.0785 \times 10^{-3} + 1.8217 \times 10^{-7} p + 4.2679 \times 10^{-12} p^2 \quad (5)$$

approximates $\rho^e$ to within less than 0.01% relative error. This implies density errors less than 0.1 kg/m³, which is an order of magnitude less than density variations expected with temperature alone. From Eq. 4, we can derive the approximate thermal expansion coefficient

$$\alpha = -\frac{1}{\rho_0} \frac{\partial \rho}{\partial T}\bigg|_p = -\frac{2C(p)\left[T - T_d(p)\right]}{\rho_0} \quad (6)$$

which is shown in Fig. 2 and changes sign at $T = T_d > T_f$ for pressures lower than $p^* = 2848$ dbar.

We note that the nonmonotonic, anomalous behavior of water density at low pressure is well known for freshwater lakes at atmospheric pressure [39] and disappears progressively with increasing salt concentration. With typical salinities of $S \lesssim 35$ g/kg, the density of Earth's oceans decreases monotonically with increasing temperatures.

## Evolution equations
The evolution equations for subglacial lakes are the Navier-Stokes equations in the Boussinesq approximation. Here, we include compressibility effects in the energy equation because we are interested in the calculation of the (small) critical heat flux at the onset of convection, but compressibility effects can otherwise be neglected when considering the (large) geothermal heat flux. In a Cartesian coordinates system $(x, y,$ and $z)$ centred on a lake's top boundary, the equations for the velocity vector $\mathbf{u}$, pressure $p$, density $\rho$, and temperature $T$ (in °C) read [40–42]

$$\rho_0 \frac{D\mathbf{u}}{Dt} + \rho_0 f \mathbf{e}_z \times \mathbf{u} = -\nabla p + \mu \nabla^2 \mathbf{u} - \rho g \mathbf{e}_z \quad (7)$$

$$\nabla \cdot \mathbf{u} = 0 \quad (8)$$

$$\rho_0 c_p \frac{DT}{Dt} - \alpha(T + T_0) \frac{Dp}{Dt} = k \nabla^2 T \quad (9)$$

where $f$ is the Coriolis frequency, $\mu$ is the dynamic viscosity, $k$ is the thermal conductivity, $c_p$ is the isobaric specific heat capacity, $g$ is the gravitational acceleration, and $T_0 = 273.15$ K; $D/Dt \equiv \partial_t + \mathbf{u} \cdot \nabla$ denotes material derivative, with $\partial_t$ the time derivative and $\nabla$ is the gradient operator. Equation 7 is the momentum equation in the Boussinesq approximation; Eq. 8 is mass conservation for an incompressible fluid; and Eq. 9 is the energy equation including pressure effects, which are relevant in the limit of small temperature variations. Note that $p$ is in pascal (Pa) in the above equations but is converted to dbar by dividing by $10^4$ when used in Eqs. 3 to 6.

We consider the Coriolis frequency at 80°S, i.e., we take $f = -1.4363 \times 10^{-4}$ rad/s; we use $\mu = 1.7 \times 10^{-3}$ kg m⁻¹ s⁻¹ and $k = 0.56$ W m⁻¹ K⁻¹, which are the dynamic viscosity and thermal conductivity values at reference pressure $p = 0$ dbar and temperature $T = 0.01$°C; we use $c_p = 4.2174 \times 10^3$ J kg⁻¹ K⁻¹; and we recall that $g = 9.81$ m/s² at and near Earth's surface. Note that $\alpha$ and $k$ may be expected to vary with $p$ and $T$. However, to the best of our knowledge, only few studies have investigated their dependence, in particular in the cold-temperature and high-pressure regimes, and reported little variations for the pressure and temperature conditions of our interest such that we take them constants [43, 44]. We denote by $\nu = \mu/\rho_0 = 1.7 \times 10^{-6}$ m²/s the constant kinematic viscosity and by $\kappa = k/\rho_0 c_p = 1.3 \times 10^{-7}$ m²/s the constant thermal diffusivity.

Subglacial lake water must be at the freezing temperature at the upper lake boundary, i.e., $T = T_f(p_i)$ at $z = 0$ m, while at the base of the lake, it is the heat flux that is enforced, i.e., $k\partial_z T = -F$ at $z = -h$, with $F > 0$ the (geothermal) heat flux and $h > 0$ the lake depth; also, $p = p_i$ at $z = 0$. Equations 7 to 9, along with the equation of state (Eq. 4), have the stationary base-state solution (denoted by overbars)

$$\overline{\mathbf{u}} = 0, \overline{T} = T_f(p_i) - \frac{zF}{k}, d_z \overline{p} = -\rho(\overline{T}, \overline{p}) g \quad (10)$$

i.e., the temperature increases linearly with depth, and the pressure is hydrostatic. For simplicity, we assume $\rho \approx \rho_0$ in the hydrostatic base-state equation, such that $\overline{p} = p_i - \rho_0 g z$ at leading order (assuming a pressure variable expressed in Pascals).

## Static stability of an ideal compressible fluid
The criterion for an ideal (dissipationless) compressible fluid with hydrostatic base-state pressure $p = p_i - \rho_0 g z$ (overbar dropped) to be locally (statically) stable is [39, 45]

$$\frac{1}{\rho_0} \frac{\partial \rho}{\partial T}\bigg|_p \frac{ds}{dz} = -\alpha\left(\frac{c_p}{T + T_0} \frac{dT}{dz} - \frac{\alpha}{\rho_0} \frac{dp}{dz}\right) = \frac{-\alpha c_p}{T + T_0}\left(\frac{dT}{dz} - \frac{dT_{ad}}{dz}\right) < 0 \quad (11)$$

where $s$ is the entropy, $T_0 = 273.15$ K (we recall that we express temperature variables in °C), and we approximate $\rho \approx \rho_0$ at leading order. $dT_{ad}/dz$ is known as the adiabatic temperature gradient and reads

$$\frac{dT_{ad}}{dz} = \frac{-\alpha(T + T_0) g}{c_p} \quad (12)$$

such that $dT_{ad}/dz > 0$ if $\alpha < 0$. Here, heating is provided from the bottom of the lake such that we always have $dT/dz < 0$. As a result, when $\alpha < 0$, equation 11 is always satisfied and the lake is stable. For $\alpha > 0$, equation 11 is not satisfied, and the lake is unstable to vertical convection if $dT/dz < dT_{ad}/dz < 0$, i.e., if the heat flux exceeds (in absolute value) the adiabatic heat flux. Thus, the two conditions for subglacial lakes experiencing geothermal heating (i.e., such that $dT/dz < 0$) to be locally unstable are

$$\alpha > 0 \tag{13}$$

$$\frac{dT}{dz} < \frac{dT_{ad}}{dz} \tag{14}$$

Note that both $\alpha$ and $dT_{ad}/dz$ are functions of $z$ when considering the base state of a subglacial lake heated from below. Specifically, $\alpha$ increases with depth, while $dT_{ad}/dz$ decreases with depth (note that it is negative and so increases in absolute value), such that equation 13 (resp. Eq. 14) is more readily satisfied at the bottom (resp. top) of the lake. As a result, it is possible to find cases where a subglacial lake is globally unstable but remains statically stable in some places. When this happens, convection is expected to occur in subregions of the water column where equations 13 and 14 are satisfied.

Equations 13 and 14 are necessary but not sufficient conditions for flow instability. The heat flux must sustain a temperature gradient with a buoyancy anomaly that is also large enough to overcome viscosity and diffusivity effects. The calculation of the exact, minimum critical heat flux leading to convection in subglacial lakes, with dissipation effects taken into account, is the result of the stability analysis described in the next section.

### Linear stability analysis

We study the stability of the base-state solution (Eq. 10) by investigating how small initial perturbations evolve over time. We expand the variables (generically represented by $X$) as

$$X = \overline{X} + X' \tag{15}$$

with primes denoting the perturbed variables. Substituting expanded variables in Eqs. 7 to 9, using Eq. 4, and linearizing, we obtain the perturbation equations

$$\rho_0 \frac{\partial \mathbf{u}'}{\partial t} + \rho_0 f \mathbf{e}_z \times \mathbf{u}' = -\nabla p' + \mu \nabla^2 \mathbf{u}' + \rho_0 \overline{\alpha} T' g \mathbf{e}_z - \overline{p}_p p' g \mathbf{e}_z \tag{16}$$

$$\nabla \cdot \mathbf{u}' = 0 \tag{17}$$

$$\rho_0 c_p (\partial_t T' - w' F/k) - \overline{\alpha}(\overline{T} + T_0)(\partial_t p' - w' \rho_0 g) = k \nabla^2 T' \tag{18}$$

where $\overline{p}_p$ is related to the small compressibility of the background state and is derived from Eq. 4 as

$$\overline{p}_p = [\rho_{11} + 2\rho_{12}\overline{p} + (\overline{T} - \overline{T}_d)^2 (C_1 + 2C_2\overline{p}) + 2\overline{C}(\overline{T} - \overline{T}_d) \\ (-T_{d1} - 2T_{d2}\overline{p})] \tag{19}$$

with subscripts 1,2 denoting the linear and leading coefficients of the polynomial expressions for $\rho_1$, $T_d$, and $C$, e.g., $T_{d1} = -2.0059 \times 10^{-3}$ K/dbar (Eqs. 3 and 5).

Since the perturbation equations do not depend explicitly on $x$, $y$, and $t$, the stability criterion can be inferred from the temporal evolution of plane waves of the form

$$X'(x, y, z, t) = \widehat{X}(z) e^{\sigma t + i(k_x x + k_y y)} + \text{c.c.}, \tag{20}$$

with $\sigma$ as the growth rate, $k_x$ and $k_y$ as the wave numbers in the $x$ and $y$ directions, and c.c. as the complex conjugate. Assuming horizontal isotropy, i.e., $k_x = k_y$, and substituting variables of the form given by Eq. 20 in Eqs. 16 to 18, we derive a one-dimensional linear eigenvalue problem for the growth rate $\sigma$, which we solve numerically with the open-source pseudospectral Dedalus code and the Eigentools package (46). We expand variables in the $z$ direction using Chebyshev modes and compute the largest growth rate $\sigma(k_\perp, F)$ as a function of wave number $k_\perp = \sqrt{k_x^2 + k_y^2}$ and heat flux $F$ for a range of input parameters ($p_i$, $h$). The critical minimum heat flux $F_c$ that destabilizes the base state is the minimum of $F$ for which there exists a wave number $k_\perp$ such that $\sigma(k_\perp, F) > 0$. We report $F_c$ in Fig. 3A. Note that the eigenvalue problem becomes challenging at large $h$, such that we limit the calculations to $h \leq 20$ m. We extrapolate to larger $h$ using scaling laws that are either asymptotically valid or conservative, i.e., such that they may overestimate $F_c$. We describe the extrapolation procedure in detail in the Supplementary Materials.

### Scaling laws for nonrotating vertical convection

We show in the Supplementary Materials that vertical convection in Antarctic subglacial lakes is better represented by nonrotating convection than geostrophic convection. As a result, we use scaling laws obtained in the idealized limit of nonrotating turbulent convection to make predictions about $\delta$, $T_{bulk}$, $U$, and $\ell$ for subglacial lakes subject to a geothermal flux $F = 50$ mW/m$^2$.

First, we remark that the conductive layer at the top of the lake includes the turbulent boundary layer and a stable layer where $\alpha < 0$ for $p_i < p^*$. In other words, we write $\delta = \delta_t + \delta_S$ with $\delta_t$ as the thickness of the turbulent boundary layer and $\delta_S$ as the thickness of the stable layer. We obtain $\delta_S$ as the positive solution of the equation $T_d(p_i + \rho_0 g \delta_S / 10^4) = T_f(p_i) + \delta_S F/k$, i.e., $\delta_S$ is found as the location $z = -\delta_S$ where the temperature of maximum density equals the temperature of the conductive base-state profile. The temperature at the base of the stable layer is correspondingly $T_S = T_f + \delta_S F/k$. The definition of the control parameter $Ra_F$ (Eq. 1) of subglacial lakes includes the effective water depth $h_{eff}$ and the characteristic thermal expansion coefficient $\alpha_{eff}$. The effective water depth is simply the region of the water column where convection occurs ($\alpha > 0$), such that we assume $h_{eff} = h - \delta_S$. Estimating $\alpha_{eff}$ inside a convective lake is difficult since the vertical profiles of temperature and $\alpha$ are unknown. Here, for simplicity, we use $\alpha_{eff} = \alpha(p_i + \rho_0 g h/10^4, T_S)$, i.e., we take $\alpha$ on the bottom boundary as the effective thermal expansion coefficient but assume that the temperature of the lake does not exceed $T_S$. This assumption is likely to underestimate $\alpha_{eff}$ but is the best conservative assumption possible without prior knowledge of the vertical temperature profile.

We use scaling laws for the Nusselt number $Nu$ and the Reynolds number $Re$ inferred as functions of $Ra$ from recent state-of-the-art numerical simulations (25, 47) to predict $\delta_t$, $T_{bulk}$, $U$, and $\ell$. We relate our flux-based Rayleigh number $Ra_F$ to $Ra$ following previous works (48, 49), i.e., such that

$$Ra_F = Ra Nu \tag{21}$$

The scaling laws for $Nu$ in (47) and $Re$ in (25) are

$$Nu = 0.16 Ra^{2/7} \tag{22}$$

$$Re = 0.18 \left(Ra - \frac{Ra}{Nu}\right)^{1/2} Pr^{-1} \tag{23}$$

Combining Eq. 1 with Eqs. 21 and 22, we can estimate $Ra = (Ra_F/0.16)^{7/9}$ and predict $\delta_t = 0.5 h_{eff}/Nu = 0.5 h_{eff}/(0.16 Ra^{2/7})$, such that

$$\delta = \frac{0.5 h_{eff}}{0.16 Ra^{2/7}} + \delta_S \qquad (24)$$

Assuming a conductive temperature profile in the stable and turbulent boundary layers, we then find

$$T_{bulk} = \frac{F\delta}{k} \qquad (25)$$

The turbulent flow velocity is inferred from $Re = U h_{eff}/\nu$ and Eqs. 22 and 23 as

$$U = \frac{\nu 0.18 \sqrt{Ra - 6.25 Ra^{5/7}}}{Pr h_{eff}} \qquad (26)$$

The estimate of the characteristic turbulent length scale, or distance between plumes, is finally obtained from equation (6.3) of reference (25) as

$$\ell = 0.8 \sqrt{RePr} (3.125 Ra^{-2/7})^{3/2} h_{eff} \qquad (27)$$

We provide a second set of predictions for $\delta_t$, $T_{bulk}$, and $U_{lsc}$ based on scaling laws inferred from the GL unifying theory of RB convection (26). We use subscript lsc for the velocity as the GL theory applies to the velocity of the large-scale circulation rather than the velocity of the plumes, as is the case in (25). The procedure for deriving $\delta_t$, $T_{bulk}$, and $U_{lsc}$ from the GL theory is the same as above, i.e., we combine Eqs. 1 and 21 with scaling laws for $Nu$ and $Re$ to derive $\delta_t = 0.5 h_{eff}/Nu$, $T_{bulk} = F\delta/k$ ($\delta_S$ is unchanged) and $U_{lsc} = \nu Re/h_{eff}$. The scaling laws for $Nu$ and $Re$ combine the expressions derived in subregions $I_u$, $III_u$, and $IV_u$ of the GL theory [see (26) and the Supplementary Materials].

## Scaling laws for rotating horizontal convection
The ice-water interface of subglacial lakes is often sloped such that a baroclinic horizontal convection flow develops along the ice-water interface. Here, we provide approximate estimates of the horizontal velocity of the baroclinic flow to compare the dynamical importance of vertical convection to horizontal convection. Besides the Prandtl number $Pr$, the control parameters for horizontal convection are the Ekman number $Ek_L$ and the Rayleigh number $Ra_L$ based on the lake's horizontal length $L$ (27, 50), i.e.

$$Ek_L = \frac{\nu}{|f| L^2}, Ra_L = \frac{g \alpha_i \Delta_i L^3}{\nu \kappa} \qquad (28)$$

where $\alpha_i$ is the effective thermal expansion coefficient near the ice ceiling, and $\Delta_i$ is the temperature difference along the tilted ice-water interface due to the changing freezing temperature $T_f(p_i)$ with the ice pressure (Eq. 2). The ice pressure drop along the ice-water interface is $\delta_{pi} = sL\rho_i g/10^4$ dbar, with $s$ the slope and $L$ in meters, such that

$$\Delta_i = T_f(p_i) - T_f(p_i + \delta_{p_i}) > 0 \qquad (29)$$

For $\alpha_i$, we take the maximum of $\alpha$ along the ice-water interface, i.e.

$$\alpha_i = \max[|\alpha(p_i, T_f(p_i))|, |\alpha(p_i + \delta_{p_i}, T_f(p_i + \delta_{p_i}))|] \qquad (30)$$

We show in the Supplementary Materials that horizontal convection in Antarctic subglacial lakes is constrained by rotation. Thus, here we use a recent scaling law for the Reynolds number $Re_{hc}$ of rotation-constrained horizontal convection (27) to estimate the characteristic velocity of the large-scale horizontal flow $V_{hc}$. For simplicity, we assume a fixed aspect ratio of $L/h = 250$, i.e., the horizontal length is 250 times the water depth, even though subglacial lakes have variable aspect ratio. We note that a ratio of 250 is on the higher end of observed aspect ratios (see Table 1), such that our estimates of the lakes' lengths may be closer to an upper than a lower bound. The scaling law for rotating horizontal convection inferred from (27) is

$$Re_{hc} = \frac{(Ra_L Ek_L)^{2/3}}{Pr} \qquad (31)$$

such that

$$V_{hc} = \frac{\nu Re_{hc}}{L} \qquad (32)$$

Note that $V_{hc} \sim L^{1/3}$, such that estimates for the horizontal velocity would be only weakly affected (weakly decreasing) by decreasing the aspect ratio.

## SUPPLEMENTARY MATERIALS
Supplementary material for this article is available at http://advances.sciencemag.org/cgi/content/full/7/8/eabc3972/DC1

**Acknowledgments:** We gratefully acknowledge fruitful discussions with K. Makinson at the British Antarctic Survey and C. Vreugdenhil at the University of Cambridge. L.-A.C. thanks T. Alboussière from ENS Lyon for the helpful discussions on the thermodynamics and stability of compressible fluids. **Funding:** This project has received funding from the European Union's Horizon 2020 Research and Innovation Programme under the Marie Skłodowska-Curie grant agreement 793450. We acknowledge PRACE for awarding us access to Marconi at CINECA, Italy. **Author contributions:** All authors conceptualized and supervised the work. L.-A.C. performed the calculations and led the writing of the original draft and revision. All authors discussed the results and reviewed and edited the paper. **Competing interests:** The authors declare that they have no competing interests. **Data and materials availability:** All data needed to evaluate the conclusions in the paper are present in the paper and/or the Supplementary Materials. All data, code, and materials will be made available on a per request basis. The simulation code Dedalus used in this work is open source.

Submitted 22 April 2020
Accepted 4 January 2021
Published 17 February 2021
10.1126/sciadv.abc3972

**Citation:** L.-A. Couston, M. Siegert, Dynamic flows create potentially habitable conditions in Antarctic subglacial lakes. *Sci. Adv.* **7**, eabc3972 (2021).

# Science Advances

**Dynamic flows create potentially habitable conditions in Antarctic subglacial lakes**

Louis-Alexandre Couston and Martin Siegert